\documentclass[twocolumn, superscriptaddress,showpacs,preprintnumbers,aps,amsmath,amssymb,amsfonts,pra]{revtex4}

\usepackage{graphicx}
\usepackage{dcolumn}
\usepackage{bm}

\begin{document}

\newcommand{\Tr}{\text{Tr }}
\newcommand{\hhf}{\bar h^{\text{HF}}}
\newcommand{\beq}{\begin{equation}}
\newcommand{\eeq}{\end{equation}}
\newcommand{\beqa}{\begin{eqnarray}}
\newcommand{\eeqa}{\end{eqnarray}}
\newcommand{\br}{\mathbf{r}}
\newcommand{\bx}{\mathbf{x}}
\newcommand{\bxp}{\mathbf{x'}}
\newcommand{\GHF}{G_{\rm{HF}}}
\newcommand{\la}{\langle}
\newcommand{\ra}{\rangle}
\newcommand{\Tra}{\textrm{Tr }}
\newcommand{\psih}{\hat{\psi}}
\newcommand{\psihd}{\hat{\psi}^{\dagger}}
\newcommand{\hpsid}{\hat{\psi}^{\dagger}}
\newcommand{\hpsi}{\hat{\psi}}
\newcommand{\hH}{\hat{H}}
\newcommand{\psin}{\Psi_0^N}
\newcommand{\psinmu}{\Psi_i^{N-1}}
\newcommand{\hU}{\hat{U}}
\newcommand{\hV}{\hat{V}}
\newcommand{\psia}{\hat{\psi}}
\newcommand{\psic}{\hat{\psi}^{\dagger}}

\newcommand{\bra}[1]{\langle #1|}l.
\newcommand{\ket}[1]{|#1\rangle}
\newcommand{\braket}[2]{\langle #1|#2\rangle}
 \arraycolsep=1.4pt

\title{Time-propagation of the Kadanoff-Baym equations for inhomogeneous systems\footnote{See J. Chem. Phys. {\bf 130} for the published version.}}
\author{Adrian Stan}
\affiliation{Department of Physics, Nanoscience Center, FIN 40014, University of Jyv\"askyl\"a,
Jyv\"askyl\"a, Finland}
\affiliation{European Theoretical Spectroscopy Facility (ETSF)}
\affiliation{Rijksuniversiteit Groningen, Zernike Institute for Advanced Materials, Nijenborgh 4, 9747AG Groningen, The Netherlands.}
\author{Nils Erik Dahlen}
\affiliation{Rijksuniversiteit Groningen, Zernike Institute for Advanced Materials, Nijenborgh 4, 9747AG Groningen, The Netherlands.}
\author{Robert van Leeuwen}
\affiliation{Department of Physics, Nanoscience Center, FIN 40014, University of Jyv\"askyl\"a,
Jyv\"askyl\"a, Finland}
\affiliation{European Theoretical Spectroscopy Facility (ETSF)}

\begin{abstract}

We have developed a time propagation scheme for the Kadanoff-Baym equations
for general inhomogeneous systems. These equations describe the time evolution of 
the nonequilibrium Green function for interacting many-body systems in
the presence of time-dependent external fields. The external fields are treated
nonperturbatively whereas the many-body interactions are incorporated perturbatively
using $\Phi$-derivable self-energy approximations that guarantee the
satisfaction of the macroscopic conservation laws of the system. 
These approximations are discussed in detail for the time-dependent Hartree-Fock,
the second Born and the $GW$ approximation.
\vspace{1.5cm}

\end{abstract}


\maketitle
\date{}

\section {Introduction} 
The recent developments in the field of molecular electronics have emphasized the need
for further development of theoretical methods that allow for a
systematic study of dynamical processes like relaxation and 
decoherence at the nanoscale. Understanding these processes is of 
utmost importance for making progress in molecular electronics, whose ultimate 
goal is to minimize the size and maximize the speed of integrated 
devices~\cite{Cuniberti:book}. To study these phenomena, theoretical methods must allow for
the possibility to study the ultrafast transient dynamics \cite{ultrafast1, ultrafast2} 
up to the picosecond \cite{pico1, pico2} and femtosecond timescale, while including 
Coulomb interactions, without violating basic conservation laws such as the continuity equation
\cite{baym62}.
A theoretical framework that incorporates these features is
the nonequilibrium Green function approach
based on the real-time propagation of the Kadanoff-Baym 
(KB) equations \cite{kadanoff62, dahlen07, danielewicz, kwong00, myohanen08, Schaefer:JOCS96,Bonitzetal:JPC96,Binderetal:PRB97}.
This method allows for systematic inclusion of electron interactions while providing results in agreement 
with the macroscopic conservation laws of the system~\cite{kadanoff62,baym62}. 
In two recent Letters~\cite{dahlen07,myohanen08} we applied the KB equations to investigate the short time dynamics
of atoms and molecules in time-dependent external fields, as well as the transport dynamics of double quantum dot devices. 
It is the aim of this paper to describe in detail the underlying method that was only briefly described
in those Letters.
This includes both a description of the theory as well as the time-propagation algorithm.
We further generalize the equilibrium method, described in two recent papers~\cite{dahlen05b,astan09},
to the nonequilibrium domain. We also extend earlier work
on the time-propagation method of the KB equations for homogeneous systems~\cite{kohler99, BonitzCompMethBook} 
to the case of inhomogeneous systems.
In the inhomogeneous case we can not take advantage of Fourier transform techniques anymore.
The KB equations become time-dependent matrix equations instead, in which the matrices are
indexed by basis function indices.
The time-stepping algorithm has to take into account the special double-time structure of
the equations which are furthermore nonlinear, inhomogeneous and non-Hermitian. Therefore,
several standard time-propagation methods can not be used.
Our approach is different from the one presented in Refs.~\cite{kohler99, BonitzCompMethBook}
by incorporating correlated initial states and the memory thereof, which is described in terms of
Green functions with mixed real and imaginary time arguments.
To simplify the time-stepping procedure, we make use of several symmetry relations of the Green function.\\
The paper is divided as follows: in section II we present the KB equations
and their symmetry properties.
In section III we discuss the conserving self-energy approximations that we use, and in section IV
we present the time-propagation method that we developed for systems described within a general basis
set representation.
Finally in section V we present a summary and conclusions.

\section{Theory}

\begin{figure}
\includegraphics[width=8.6cm]{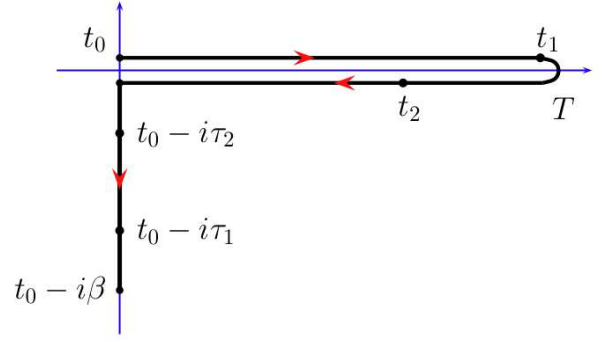}
\caption{Keldysh contour.
The depicted contour allows for the calculation of observables
for times $t_0 \leq t \leq T$. The initial Green function is calculated on the imaginary track
$[t_0,t_0-i\beta]$.
As we propagate the KB equations in time, for real times $t > t_0$,
the turning point of the time-contour at $t=T$ moves to the right along the real time axis.}
\label{fig:contour}
\end{figure}

We consider a many-body system that is initially 
in equilibrium at a temperature $T$ and with a chemical potential $\mu$. 
At an initial time $t_0$ the system is exposed to a time-dependent external field.
This external field can, for instance, be a bias voltage in a quantum transport
case, or a laser pulse. The field forces the system out of equilibrium and
we aim to describe the time-evolution of this nonequilibrium state. 
In second quantization the time-dependent Hamiltonian of the system reads
(throughout this paper we use atomic units $\hbar=m=e=1$)
\beqa
&&\hat{H}(t)=\int\; d\mathbf{x}\;\hat{\psi}^\dagger(\mathbf{x}) h(\mathbf{x},t) \hat{\psi}(\mathbf{x}) + \nonumber \\
&&+\frac{1}{2}\int\int\; d\mathbf{x_1}  d\mathbf{x_2} \hat{\psi}^\dagger(\mathbf{x_1})\hat{\psi}^\dagger(\mathbf{x_2})v(\br_1,\br_2)\hat{\psi}(\mathbf{x_2})\hat{\psi}(\mathbf{x_1}), 
\eeqa 
where $\bx=(\br,\sigma)$ denotes the space- and spin coordinates. The two-body interaction will, in general, be a Coulombic repulsion 
of the form $v(\br_1,\br_2)=1/|\mathbf{r_1}-\mathbf{r_2}|$. 
The one-body part of the Hamiltonian is
\beq
h(\bx,t)=-\frac{1}{2}\nabla^2+w(\bx,t)-\mu,
\label{eq:singleparticleH}
\eeq
where $w(\bx, t)$ is a time-dependent external potential.
The chemical potential $\mu$ of the initial equilibrium system is absorbed in the 
one-body part of the Hamiltonian.
The expectation value of an operator $\hat O$, for a system initially in
thermodynamic equilibrium ($t < t_0$), is given by
\beq
\langle \hat O \rangle = \Tra \{ \hat \rho \hat O \},
\eeq
where $\hat \rho = e^{-\beta \hH_0} / \Tra { e^{-\beta \hH_0}}$ is the statistical operator, $\hH_0$ 
is the time-independent Hamiltonian that describes the system before the time-dependent field is applied and 
$\beta=1/k_B T$ is the inverse temperature. The trace here represents a summation over a complete set of 
states in Fock space~\cite{fetterwalecka}. After the time-dependent external field is switched on at time $t_0$, 
the expectation value is given by
\beq
\langle \hat O (t) \rangle =  \frac{\text{Tr}\{\hat{U}(t_0-i\beta,t_0) \hat O_H(t)\}}{\Tra \{ \hat{U}(t_0-i\beta,t_0)\}},
\label{eq:expval1}
\eeq
where $\hat O_H(t) = \hU(t_0, t) \hat O \hU(t, t_0)$ is the operator $\hat O$ in the Heisenberg picture
and $\hU (t_2,t_1)=T[ \exp(-i \int_{t_1}^{t_2} dt \hH (t)) ]$, for $t_2 > t_1$,  is the time-ordered 
evolution operator of the system.
We further wrote $\exp(-\beta  \hat{H}_0)=\hat{U}(t_0-i\beta,t_0)$ as an evolution operator in imaginary time.
If we read the time arguments in Eq.(\ref{eq:expval1}) from right to left we see that they follow a
time-contour as displayed in Fig.\ref{fig:contour}. This contour is also known as the
Keldysh contour~\cite{keldysh65,lnop}. A more detailed inspection of Eq.(\ref{eq:expval1})
then shows that the expectation value can also be written as a contour-ordered 
product~\cite{lnop,stefanucci_almbladh2004, wagner91, dahlenKiel324, dahlenKiel340}. 
The one-particle Green function is then defined as a countour-ordered product of a creation and
an annihilation operator
\beqa
G(1,2) = -i \la T_C [\psia_H(1) \psic_H(2) ] \ra,
\label{eq:gdef} 
\eeqa
where $T_C$ denotes the time-ordering operator on the contour and 
where we used the compact notation $1=(\bx_1, t_1)$ and $2=(\bx_2, t_2)$.
If we consider the Green function at time $t_1=t_0-i\beta$ and use the cyclic property of the trace, we 
find that $G(\bx_1 t_0-i\beta,2)=-G(\bx_1 t_0,2)$~\cite{dahlenKiel324}. 
Hence, the Green function defined in Eq.~(\ref{eq:gdef}) obeys the boundary conditions
\beqa
G(\bx_1 t_0,2)&=&-G(\bx_1t_0-i\beta,2), \\ 
G(1,\bx_2 t_0)&=&-G(1,\bx_2 t_0-i\beta).
\eeqa 
The Green function satisfies the equation of motion
\beq
[i\partial_{t_1}- h(1)]G(1,2)=\delta(1,2)+\int_C d3 \Sigma(1,3)G(3,2),
\label{eq:motion}
\eeq
as well as a corresponding adjoint equation~\cite{danielewicz,dahlenKiel324}.
In Eq.(\ref{eq:motion}) the time-integration is carried out along the contour $C$.
The self-energy $\Sigma$ incorporates the effects of exchange and correlation in 
many-particle systems and is a functional of the Green function
that can be defined diagrammatically~\cite{danielewicz,fetterwalecka}.
The Green function can be written as
\beqa
G(1,2)
&=&\theta(t,t')G^>(1,2)+\theta(t',t)G^<(1,2),
\label{eq:Ggtrless}
\eeqa
where $\theta$ is a step function generalized to arguments on 
the contour {\it i.e.} with 
$\theta(t,t')=1$ if $t$ is later on the contour than $t'$ and zero otherwise~\cite{danielewicz}. 
The greater and lesser components $G^>$ and $G^<$ respectively, have the explicit form
\beqa
G^>(1,2)&=&-i \la\hat{\psi}_H(1)\hat{\psi}^\dagger_H(2) \ra,\\
G^<(1,2)&=&i \la\hat{\psi}^\dagger_H(2)\hat{\psi}_H(1)\ra.
\eeqa
When one of the arguments is on the vertical track of the contour, we adopt the  
notation~\cite{stefanucci_almbladh2004}
\beqa
G^{\rceil} (1 , \bx_2,-i\tau_2) &=& G^< (1, \bx_2,t_0-i\tau_2), \\
G^{\lceil} (\bx_1,-i\tau_1, 2) &=& G^> (\bx_1,t_0-i\tau_1, 2).  
\eeqa
Finally, for the case when both time 
arguments are on the imaginary track of the contour, we have the so-called Matsubara Green function $iG^M$~\cite{fetterwalecka}
\beqa
iG^M (\bx_1\tau_1, \bx_2\tau_2)&&=G(\bx_1t_0-i\tau_1,\bx_2t_0-i\tau_2), 
\label{eq:Gmatsubara}
\eeqa 
which is a well-known object from the equilibrium theory. The factor $i$ in the definition of Eq.(\ref{eq:Gmatsubara})
is a convention which ensures that $G^M$ is a real function.
The self-energy $\Sigma$ has a similar general structure as the Green function 
\beq
\Sigma(1,2)=\Sigma^{HF}(1,2)+\theta(t, t')\Sigma^>(1,2)+\theta(t',t)\Sigma^<(1,2).
\label{eq:selfenergyoperator}
\eeq
The main difference with Eq.(\ref{eq:Ggtrless}) is the appearance of the term $\Sigma^{HF}$ which is
proportional to a contour delta function $\delta(t_1,t_2)$
in the time coordinates~\cite{danielewicz}. This term has the explicit form
\beqa
\Sigma^{HF} [G](1,2) &=& \delta(t_1,t_2) \Sigma^{HF}(\bx_1,\bx_2,t_1),
\label{eq:sigmaHF}
\eeqa
where
\beqa
\lefteqn{ \Sigma^{HF}(\bx_1,\bx_2,t) =  
i G^<(\bx_1 t,\bx_2 t) v(\bx_1,\bx_2) } \nonumber \\
 && - i \delta (\bx_1-\bx_2) \int d\bx_3 v(\bx_1,\bx_3) G^< (\bx_3 t,\bx_3 t).
\label{eq:sigHF}
\eeqa
The structure of this self-energy is that of the Hartree-Fock (HF) approximation.
However, in general we will evaluate this expression for Green functions $G$ 
obtained beyond HF level (see section~\ref{sec:self}).
Using the form of the self-energy of Eq.(\ref{eq:selfenergyoperator}) 
the contour integrations can be readily carried out~\cite{danielewicz,lnop}
and we find separate equations for the different Green functions $G^\lessgtr,G^{\rceil \lceil}$ and $G^M$.
To display their temporal structure more clearly we surpress the spatial indices 
of the Green functions and self-energies. Alternatively, these quantities may be regarded
as matrices~\cite{dahlen07}. 
On the imaginary track of the contour we obtain
\beq
[-\partial_\tau-h]G^M(\tau)=\delta(\tau)+\int_{0}^{\beta} d\bar{\tau}\Sigma^M(\tau-\bar{\tau})G^M(\bar{\tau}),
\label{eq:motionM}
\eeq
where the Green function and the self-energy are functions of the time-differences only, {\em i.e.}
$iG^M (\tau_1-\tau_2)=G(-i\tau_1, -i\tau_2)$ and $i\Sigma^M (\tau_1-\tau_2)=\Sigma(-i\tau_1, -i\tau_2)$, since
the Hamiltonian is time-independent (and equal to $\hH_0$) on the imaginary track.
Equation~(\ref{eq:motionM}), which determines the Green function of the equilibrium system,
has been treated in detail in references~\cite{dahlen05b,astan09}.
For the other Green functions we obtain
\beqa
i\partial_t G^\lessgtr(t,t')&=&h^{HF}(t)G^\lessgtr (t, t')+I^\lessgtr_1(t,t'), 
\label{kbi1} \\
-i\partial_{t'} G^\lessgtr(t,t')&=&G^\lessgtr (t,t') h^{HF}(t')+I^\lessgtr_2(t,t'),
\label{kbi2} \\
i\partial_t G^\rceil(t,-i\tau)&=&h^{HF}(t) G^\rceil(t,-i\tau)+I^\rceil(t,-i\tau),
\label{kbi3}\\
-i\partial_t G^\lceil(-i\tau,t)&=&G^\lceil(-i\tau,t) h^{HF}(t)+I^\lceil(-i\tau,t),
\label{kbi4}
\eeqa
where $h^{HF}(t)=h(t) + \Sigma^{HF}(t)$ and $\Sigma^{HF}(t)$ is
given by Eq.(\ref{eq:sigHF}).
The retarded and advanced functions for $G$ and $\Sigma$ are defined according to 
\beq
F^{R/A}(t,t')=\pm \theta(\pm t \mp t') [F^{>}(t,t')-F^{<}(t,t')],
\label{retadv}
\eeq
with $F$ replaced by $G$ and $\Sigma$ respectively.
The so-called collision terms $I^\lessgtr$ and $I^{\rceil  \lceil}$ have the form 
\beqa
I^\lessgtr_1(t,t')&=&\int_0^{t} d\bar t \Sigma^R(t,\bar t) G^\lessgtr(\bar t, t') +  \int_0^{t'} d\bar t \Sigma^\lessgtr (t, \bar t) G^A(\bar t, t')\nonumber \\ 
 &+& \frac{1}{i} \int_0^{\beta} d\bar \tau \, \Sigma^\rceil(t, -i \bar \tau) G^\lceil(-i \bar \tau, t'), 
\label{i1} \\
I^\lessgtr_2(t,t')&=&\int_0^{t} d\bar t  G^R(t,\bar t) \Sigma^\lessgtr(\bar t, t') +  \int_0^{t'} d\bar t G^\lessgtr (t, \bar t) \Sigma^A(\bar t, t')   \nonumber \\ 
 &+& \frac{1}{i} \int_0^{\beta} d\bar \tau \, G^\rceil(t, -i \bar \tau) \Sigma^\lceil(-i \bar \tau, t'), 
 \label{i2}\\
I^\rceil(t,-i\tau) &=& \int_0^t d\bar t \,\Sigma^R(t,\bar t) G^\rceil(\bar t, -i \tau) \nonumber \\ 
&+& \int_0^{\beta} d\bar \tau \, \Sigma^\rceil (t, -i \bar \tau) G^M(\bar \tau- \tau),
\label{i3} \\
I^\lceil(-i\tau, t) &=& \int_0^t d\bar t \, G^\lceil(-i \tau, \bar t ) \Sigma^A(\bar t,t) \nonumber \\ 
&+& \int_0^{\beta} d\bar \tau \, G^M( \tau- \bar \tau)  \Sigma^\lceil ( -i \bar \tau, t).
\label{i4}
\eeqa
These equations are readily derived using the conversion table of Ref.~\cite{lnop}.
From the symmetry relations 
\beqa
G^{\lessgtr}(t,t')^\dagger &=& -G^{\lessgtr}(t',t), 
\label{eq:symG}\\
\Sigma^{\lessgtr}(t,t')^\dagger &=& -\Sigma^{\lessgtr}(t',t), 
\label{eq:symSig}
\eeqa
it follows that we only need to calculate $G^>(t,t')$ and $\Sigma^>(t,t')$ for $t>t'$ and 
$G^<(t,t')$ and $\Sigma^<(t,t')$ for $t\leq t'$. 
These equations imply that $I_{1,2}^\lessgtr (t,t') = - I_{2,1}^\lessgtr (t',t)^\dagger$.
We further have 
\beqa
G^\lceil (-i \tau,t) &=& G^\rceil (t, -i (\beta-\tau))^\dagger, 
\label{eq:Glrsym}\\
\Sigma^\lceil (-i \tau,t) &=& \Sigma^\rceil (t, -i (\beta-\tau))^\dagger.
\label{eq:Sigmalrsym}
\eeqa
The symmetry relations (\ref{eq:symG}) and (\ref{eq:Glrsym}) for the Green function follow directly from 
its definition, whereas
the symmetry relations (\ref{eq:symSig}) and (\ref{eq:Sigmalrsym}) for the self-energy 
follow from Eqs.(3.19) and (3.20) of Ref.~\cite{danielewicz}.
Another consequence of equations (\ref{eq:Glrsym}) and (\ref{eq:Sigmalrsym}) is that $I^\lceil(-i\tau,t)=\left[I^\rceil(t,-i(\beta-\tau))\right]^\dagger$,
which means that in practice it is sufficient to calculate only $I_1^>,I_2^<$ and $I^\rceil$.
Eqs.(\ref{kbi1}) to (\ref{kbi4}) are known as the Kadanoff-Baym 
equations~\cite{kadanoff62,danielewicz}.\\
Once the Matsubara Green function $G^M(\tau)$ is obtained from Eq.(\ref{eq:motionM}),
the Green functions $G^x (x=\lessgtr, \rceil \lceil )$ can be calculated by time propagation. 
Their initial conditions are
\beqa
G^>(0,0)&=&iG^M(0^+),
\label{igg}\\
G^<(0,0)&=&iG^M(0^-),
\label{igl}\\
G^\rceil(0,-i\tau)&=&iG^M(-\tau),
\label{gl}\\
G^\lceil(-i\tau,0)&=&iG^M(\tau).
\label{gr}
\eeqa
The KB equations, together with the initial conditions, completely determine the Green functions
for all times once a choice for the self-energy has been made. The form of the self-energy will be the topic
of the next section.

\section{Self-energy approximations} 
\label{sec:self}

In the applications of the KB equations it is possible to guarantee that the
macroscopic conservation laws, such as those of particle, momentum and energy conservation, are obeyed.
Baym~\cite{baym62} has shown that this is the case whenever
the self-energy is obtained from a functional $\Phi[G]$, such that
\beq
\Sigma(1, 2)=\frac{\delta \Phi}{\delta G(2, 1)}.
\label{eq:sigcons}
\eeq
Such approximations to the self-energy are called conserving 
or $\Phi$-derivable approximations. 
Well-known conserving approximations are the Hartree-Fock, the second Born~\cite{kadanoff62}, the $GW$~\cite{hedin65},
and the $T$-matrix~\cite{kadanoff62} approximation.
In our work we implemented the first three of these.\\
{\em The second Born approximation --} 
This
 approximation for the self-energy
consists of the two diagrams to second order in the two-particle interaction~\cite{kadanoff62,baym61}
 \beq
 \Sigma(1,2)= \Sigma^{HF}(1,2)+\Sigma^{(2)}(1,2),
 \label{eq:secondborn}
 \eeq
 where $\Sigma^{HF}$ is the HF part of the self energy of Eq.(\ref{eq:sigmaHF})
and $\Sigma^{(2)}=\Sigma^{(2a)}+\Sigma^{(2b)}$ is the sum of the two terms
 \beqa
&\Sigma^{(2a)}&(1,2)=-i^2 G(1, 2) \int d3\,d4\, v(1,3)\nonumber\\
&&\times G(3,4)G(4,3) v(4,2),
 \label{eq:secondborndirect} \\
&\Sigma^{(2b)}&(1, 2)= i^2\int d3\,d4\, G(1, 3)v(1, 4)G(3, 4) \nonumber\\
&&\times G(4, 2)v(3, 2),
 \label{eq:secondbornexchange}
 \eeqa
where $v(1,2)=v(\bx_1,\bx_2) \delta(t_1,t_2)$.
These terms are usually referred to as the second-order direct and exchange terms.
This approximation to the self-energy has been discussed in detail for the equilibrium
case in Ref.~\cite{dahlen05b}. For the nonequilibrium case we need to calculate
the various components $\Sigma^x (x=\lessgtr,\rceil \lceil)$.
These are explicitly given by
\beqa
&\Sigma^{(2a),\lessgtr }&(1,2)=-i^2 G^\lessgtr (1, 2) \int d3\,d4\,  v(1,3)\nonumber\\
&&\times G^\lessgtr(3,4)G^\gtrless (4,3) v(4,2),\\
&\Sigma^{(2a),\rceil \lceil}&(1,2)=-i^2\int d3\,d4\, G^{\rceil \lceil}(1, 2) v(1,3)\nonumber\\
&&\times G^{\rceil \lceil}(3,4)G^{\lceil \rceil}(4,3) v(4,2),
\eeqa
for the direct diagram, and 
\beqa
&\Sigma^{(2b),\lessgtr}&(1, 2)=i^2\int d3\,d4\, G^\lessgtr (1, 3)v(1, 4)G^\gtrless (3, 4) \nonumber\\
&&\times G^\lessgtr (4, 2)v(3, 2), \\
&\Sigma^{(2b),\rceil \lceil}&(1, 2)=i^2\int d3\,d4\, G^{\rceil \lceil} (1, 3)v(1, 4)
G^{\lceil \rceil} (3, 4) \nonumber\\
&&\times G^{\rceil \lceil} (4, 2)v(3, 2), 
 \eeqa
for the second-order exchange diagram.
These expressions follow immediately from Eqs.(\ref{eq:secondborndirect}) and (\ref{eq:secondbornexchange}) 
with help of the conversion table of Ref.~\cite{lnop}.\\
{\em The $GW$ approximation --}
In the $GW$ approximation the exchange-correlation part of the self-energy
is given as a product of the Green function $G$ with a 
 dynamically screened
interaction $W$~\cite{hedin65} .  
The screened interaction $W$ satisfies the equation
\begin{eqnarray}
&&W(1, 2)= v(1, 2)+\int d3 d4 v(1,3)P(3, 4)W(4, 2).   
\label{eq:w}
\end{eqnarray}
Here, $v$ is the bare Coulomb interaction, and 
\beq
P(1, 2)=-iG(1,2)G(2,1),
\label{eq:p}
\eeq
is the irreducible polarization~\cite{hedin65}.
However, since the first term in Eq.(\ref{eq:w})
is singular in time (proportional to a delta function)
it is convenient, for numerical purposes, to define its time-nonlocal
part $\tilde{W}=W-v$~\cite{astan09}.
From Eq.(\ref{eq:w}) it follows that
\beqa
\tilde{W}(1,2) &=& \int d3 d4 v(1,3) P(3,4) v(4,2) \nonumber \\
&& + \int d3 d4 v(1,3) P(3,4) \tilde{W}(4,2).
\label{eq:Wtilde}
\eeqa
In terms of $\tilde{W}$, the self-energy has the form~\cite{hedin65}
\beq
\Sigma (1, 2)=\Sigma^{HF}(1,2) + iG (1, 2) \tilde{W} (1, 2).
\label{eq:gw}
\eeq
The part $\Sigma_{\textrm{c}}=i G \tilde{W}$
represents the correlation part of the self-energy
and has the components
\beqa
\Sigma_{\textrm{c}}^{\lessgtr}(1,2)&=&iG^{\lessgtr}(1,2)\tilde{W}^{\gtrless}(2,1),
\label{sl}\\
\Sigma_{\textrm{c}}^{\rceil\lceil}(1,2)&=&iG^{\rceil\lceil}(1,2)\tilde{W}^{\lceil\rceil}(2,1).
\label{src}
\eeqa
From the fact that $\tilde{W}(1,2)$ has the same symmetries as
the contour-ordered density response function~\cite{hedin65}
$\chi (1,2) = - i \langle T_C [\hat{n}_H (1) \hat{n}_H (2)] \rangle$,
where $\hat{n}$ is the density operator, it follows that
\beqa
\tilde{W}^\lessgtr (2,1) &=& \tilde{W}^\gtrless (1,2)=-[\tilde{W}^\gtrless (2,1)]^*, 
\label{eq:Wsym1}\\
W^\rceil(1,2) &=& W^\lceil(2,1).
\label{eq:Wsym2}
\eeqa
In the following, we will again surpress the spatial coordinates in order to
display the temporal structure of the equations more clearly.
From the symmetry relations (\ref{eq:Wsym1}), (\ref{eq:Wsym2}), (\ref{sl}) and (\ref{src}), and the 
fact that we only need $\Sigma^> (t,t')$ for $t > t'$ and
$\Sigma^< (t,t')$ for $t \leq t'$, it follows that we only need to 
calculate $\tilde{W}^\rceil (t,-i\tau)$, and $\tilde{W}^< (t,t')$
for $t \leq t'$. The latter obey the equations:
\beqa
\tilde{W}^<(t,t')&=&vP^<(t,t')v+vX^< (t,t'),
\label{w1}\\
\tilde{W}^\rceil(t,-i\tau)&=&vP^\rceil(t,-i\tau)v+vX^\rceil (t,-i\tau),
\label{w4}
\eeqa
where
\beqa
P^< (t,t') &=& -i G^< (t,t') G^> (t',t), \\
P^\rceil (t,-i\tau) &=& -i G^\rceil (t,-i\tau) G^\lceil (-i\tau,t),
\eeqa
and where the terms $X^<$ and $X^\rceil$ are given by
\beqa
X^<(t,t')&&=\int_0^{t'}d\bar{t}  P^<(t,\bar{t})\tilde{W}^A(\bar{t}, t')\nonumber \\
&&+\int_0^t d\bar{t} P^R(t,\bar{t})\tilde{W}^<(\bar{t},t')\nonumber \\
&&+\int_0^{\beta}d\bar{\tau} P^\rceil(t,-i\bar{\tau})\tilde{W}^\lceil(-i\bar{\tau},t'),
\label{eq:x1}\\
X^\rceil(t,-i\tau)&&=\int_0^{t'} d\bar{t} P^R(t,\bar{t})\tilde{W}^\rceil(\bar{t}, -i\tau) \nonumber \\
&&+\int_0^{\beta}d\bar{\tau} P^\rceil(t,-i\bar{\tau})\tilde{W}^M(\bar{\tau}-\tau),
\label{eq:x4}
\eeqa
with the retarded and advanced quantities defined as in Eq.(\ref{retadv}).
The initial conditions for $\tilde{W}^<$ and $\tilde{W}^\rceil$ are
 \beqa
\tilde{W}^<(0,0)&=&i\tilde{W}^M(0^-),
\label{iwl}\\
\tilde{W}^\rceil(0,-i\tau)&=&i\tilde{W}^M(-\tau),
\label{wl}
\eeqa
where $i\tilde{W}^M (\tau-\tau')=\tilde{W}(t_0-i\tau, t_0-i\tau')$ is the Matsubara interaction
discussed in detail in Ref.~\cite{astan09}.

\section{Time-propagation of the Kadanoff-Baym equations}

In the following, we will describe the time-propagation method which we employed to
solve the KB equations.
This method can be applied to general Hamiltonians 
containing one- and two-body interactions,
and is further independent of the explicit form of the self-energy.\\
\begin{figure}
    \includegraphics[width=8.6cm]{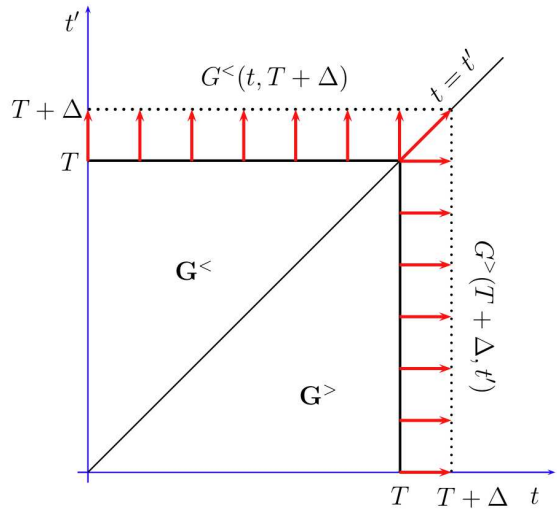}
    \caption{Time-stepping in the $(t,t')$-plane. $G^>(t,t')$ is calculated for $t > t'$
and $G^<(t,t')$ is calculated for $t \leq t'$.}
    \label{fig:timesquare}
\end{figure}
The time-propagation method is applied to the KB equations in matrix form.
This matrix form is obtained by expressing
the Green function in terms of a set of basis functions $\phi_i (\bx)$,
which we choose to be Hartree-Fock orbitals~\cite{dahlen05b,dahlen07,astan09}
\beq
G(\bx t,\bx' t')=\sum_{ij} G_{ij}(t,t') \phi_i(\bx) \phi_j^*(\bx').
\label{eq:Greenbasis}
\eeq 
When Eq.(\ref{eq:Greenbasis}) is inserted in the expressions for the 
self-energy we obtain a basis set representation of the self-energy
involving the matrices $G_{ij}(t,t')$ and the two-electron integrals
which are given as integrals of orbital products with the two-body interaction $v$.
All the quantities therefore become time-dependent matrices and all
products are to be interpreted as matrix products.
We will, however, surpress all matrix indices
to display the temporal structure of the equations more clearly.
Explicit expressions of the matrix form of the second Born and $GW$ self-energy
are given in Refs.\cite{dahlen05b,dahlen07,astan09}.\\
We start by discussing the time-propagation of $G^>$ and $G^<$.
Due to the symmetry relations Eq.(\ref{eq:symG}) and (\ref{eq:symSig})
we only need to calculate $G^>(t,t')$ for $t>t'$ and 
$G^<(t,t')$ for $t\leq t'$. 
From Eqs.(\ref{kbi1}) and (\ref{kbi2}) it then follows that 
$G^>$ must be time-stepped in the first time-argument and $G^<$
in the second one. We thus need to calculate
$G^> (T+\Delta,t')$  and $G^< (t,T+\Delta)$ for a
small time step $\Delta$, 
from the knowledge of $G^\gtrless (t,t')$ for $t,t' \leq T$. 
The symmetry relations (\ref{eq:symG}) then immediately provide us
with $G^> (t',T+\Delta)$ and $G^< (T+\Delta,t)$ as well.
The time-stepping procedure is illustrated in Fig.\ref{fig:timesquare}
that displays the $(t,t')$-plane, in which at a given time $T$ all the quantities
inside the square with sides equal to $T$, are known.
The time-step $G^<(t,T) \rightarrow G^< (t,T+\Delta)$
corresponds to a shift of the upper side of the time square with $\Delta$
{\em i.e.} a shift from the solid to the dotted line in Fig.\ref{fig:timesquare}.
Similarly the time-step $G^>(T,t') \rightarrow G^> (T+\Delta,t')$ corresponds
to a shift of the righthand side of the time square with $\Delta$.
We further need to make a step $G^< (T,T) \rightarrow G^< (T+\Delta,T+\Delta)$ along the
time diagonal $t=t'$. 
The propagation of $G^\lceil(-i\tau,t)$ and $G^\rceil (t,-i\tau)$
requires a time-step in the real time coordinate $t$ for fixed imaginary time points $\tau$.\\
Note that the righthand sides of Eqs.(\ref{kbi1}) to (\ref{kbi4}) 
depend on the Green functions at the times $T+\Delta$, which are not
known at time $T$.
We therefore carry out the time-step $T \rightarrow T+\Delta$ twice.
After taking the time step for the first time, we recalculate the righthand sides of Eqs.(\ref{kbi1}) to (\ref{kbi4})
and repeat the time-step $T \rightarrow T+\Delta$ using an average of the old and new collision and HF terms.
Since the term $h^{HF}(t)$ in Eqs.(\ref{kbi1}) to (\ref{kbi4}) 
can attain large values, it is favorable to eliminate this term from the time-stepping equations. 
For each time-step $T \rightarrow T+\Delta$ we therefore absorb the term in a time-evolution 
operator of the form
\beq
U(t) = e^{-i \hhf (T) t},
\label{eq:HFevolution}
\eeq
where $\bar{h}^{HF}(T)=h(T+ \Delta/2)+ \Sigma^{HF} (T)$, where
$h$ is the one-body part of the Hamiltonian of Eq.(\ref{eq:singleparticleH}).
The one-body Hamiltonian $h(t)$ is explicitly known as a function of time and
can be evaluated at half the time-step. The term $\Sigma^{HF}$ is only known
at time $T$ and will be recalculated in the repeated time-step.
In terms of the operator $U(t)$ of (\ref{eq:HFevolution}) we
define new Green function matrices $g^x (x=\lessgtr,\rceil \lceil)$, as
\beqa
G^\lessgtr(t_1,t_2)&=&U(t_1) g^\lessgtr(t_1, t_2) U^\dagger(t_2),
\label{gm1}\\
G^\rceil(t_1,-i\tau_2)&=&U(t_1) g^\rceil(t_1,-i\tau_2),
\label{gm2}\\
G^\lceil(-i\tau_1,t_2)&=&g^\rceil(-i\tau_1,t_2)  U^\dagger(t_2).
\label{gm3}
\eeqa
We can now transform Eqs.(\ref{kbi1}) to (\ref{kbi4}) into equations
for $g^x$.
For instance, $g^>$ satisfies the equation
\beqa
i \partial_{t} g^> (t,t') &=& U^\dagger (t) (h^{HF}(t)-\bar{h}^{HF}) G^> (t,t') U (t') \nonumber \\
&& + U^\dagger (t) I_1^> (t,t') U (t').
\label{eq:Ggrt}
\eeqa
Since $\bar{h}^{HF} \approx h^{HF}(t)$ for times $T \leq t \leq T + \Delta$, we can neglect
for these times 
the first term on the right hand side of Eq.(\ref{eq:Ggrt}). We then find
\beqa
&&G^>(T+\Delta,t_2)=U(T+\Delta) g^>(T+\Delta, t_2) U^\dagger(t_2)=\nonumber \\
&=&U(T+\Delta) \left[ g^>(T, t_2) + \int_T^{T+\Delta} dt \partial_t g^>(t, t_2)\right] U^\dagger(t_2)\nonumber\\
&\approx& U(\Delta) G^>(T, t_2) - \nonumber \\
&& i U(\Delta) \left\{ \int_{T}^{T+\Delta} dt\, e^{i\bar h^{\text{HF}}(t-T)} \right\} I^>_1(T,t_2)\nonumber\\
&=& U(\Delta) G^>(T, t_2) - V(\Delta)  I^>_1(T,t_2),
\label{eq:stepGgrt},
\eeqa
where $V(\Delta)$ is defined as
\beq
V(\Delta)=(\bar h^{\text{HF}})^{-1}[ 1-e^{-i\bar h^{\text{HF}}\Delta}].
\eeq
Similarly for $G^<$, which is propagated using Eq.(\ref{kbi2}), we find the equation
\beqa
G^<(t_1,T+\Delta) &=&
G^<(t_1, T)  U^\dagger(\Delta) \nonumber \\
&& -  I_2^<(t_1,T) V^\dagger(\Delta).
\label{eq:stepGless}
\eeqa
For time-stepping along the time-diagonal we use 
\beqa
i \partial_t G^< (t,t) &=& [ h^{HF} (t),  G^< (t,t) ] \nonumber \\
&& + I_1^< (t,t) -I_2^< (t,t),
\eeqa
which follows directly from a combination of the equations for
$G^<$ of Eqs.(\ref{kbi1}) and (\ref{kbi2}).
The corresponding equation for $g^< (t,t)$ on the time diagonal then becomes
\beqa
i \partial_t g^< (t,t) &=& U^\dagger(t) [ h^{HF} (t)-\bar{h}^{HF},  G^< (t,t) ] U(t) \nonumber \\
&&+ U^\dagger(t) (I_1^< (t,t) -I_2^< (t,t)) U(t).
\eeqa
From this equation we then obtain
\beqa
&&G^<(T+\Delta, T+\Delta)= \nonumber \\
&=&U(T+\Delta) g^<(T+\Delta, T+\Delta) U^\dagger(T+\Delta)\nonumber\\
&=& U(\Delta) G^<(T,T) U^\dagger(\Delta) - \nonumber \\
&&i U(\Delta) \left[ \int_0^{\Delta} dt \, U^\dagger(t)  I_{12} U(t)\right] U^\dagger(\Delta),
\label{eq:stepGdiag}
\eeqa
where we defined $I_{12}=I_1^< (T,T)- I_2^< (T,T)$.
By using the operator expansion 
\beqa
e^ABe^{-A} &=& B + [A,B] + \frac{1}{2} [A, [A, B]] \nonumber \\
&& + \frac{1}{3} [A, \frac{1}{2} [A, [A, B]]] + \ldots,
\eeqa
it follows that
\beq
-i \int_0^{\Delta} dt\, U^\dagger(t) I_{12} U(t) =\sum_{n=0}^{\infty}C^{(n)},
\label{eq:I12integral}
\eeq
where 
\beq
C^{(n)} = \frac{i \Delta}{n+1} [ \bar{h}^{HF} , C^{(n-1)} ],
\eeq
and $C^{(0)}=-i \Delta I_{12}$. If we insert Eq.(\ref{eq:I12integral}) into
Eq.(\ref{eq:stepGdiag}) we finally obtain
\beq
G^<(T+\Delta, T+\Delta) =  U(\Delta)\left[ G^<(T,T) + \sum_{n=0}^{\infty}C^{(n)} \right]
U^\dagger(\Delta)
\eeq
We found that keeping terms for $n \leq 3$ only, yields sufficient accuracy. 
We now consider the time propagation for the mixed real and imaginary time Green
functions.
For $g^\rceil$ we have the equation
\beqa
i \partial_t g^\rceil (t,-i\tau) &=& U(t)^\dagger (h^{HF}(t) - \bar{h}^{HF}) G^\rceil (t,-i \tau) \nonumber \\ 
&& +  U(t)^\dagger I^\rceil (t,-i\tau).
\eeqa
This yields, similarly as in Eq.(\ref{eq:stepGgrt}) and (\ref{eq:stepGless})
\beqa
G^\rceil(T+\Delta,-i\tau_2) 
&=& U (\Delta) G^\rceil(T,-i\tau_2) \nonumber \\
&& - V(\Delta) I^\rceil(T, -i\tau_2).
\label{eq:stepGr}
\eeqa
Finally, for $G^\lceil$ we have
\beqa
G^\lceil(-i\tau_1, T+\Delta)
&=& G^\lceil(-i\tau_1, T ) U(\Delta)^\dagger \nonumber \\ 
&& - I^\lceil (-i\tau_1, T) V(\Delta)^\dagger.
\label{eq:stepGl}
\eeqa
The Eqs. (\ref{eq:stepGgrt}), (\ref{eq:stepGless}), (\ref{eq:stepGdiag}), 
(\ref{eq:stepGr}) and (\ref{eq:stepGl}) form the basis of the time-stepping algorithm.
At each time-step, it requires  the construction of the step operators $U(\Delta)$ and $V(\Delta)$
and therefore the diagonalization of $\bar{h}^{HF}$ for every time-step.
As mentioned before, the righthand sides of Eqs.(\ref{kbi1}) to (\ref{kbi4}) 
depend on the Green functions at the times $T+\Delta$ which are not
known at time $T$. We therefore carry out the time-step $T \rightarrow T+\Delta$ twice. 
The procedure is as follows:\\
\\
(1) The collision integrals and $\bar{h}^{HF}$ at time $T$ are calculated from the Green functions
 for times $t,t' \leq T$. \\ 
(2) A step in the Green function $G(T)\rightarrow G(T+\Delta)$ is taken
according to Eqs.(\ref{eq:stepGgrt}), (\ref{eq:stepGless}), (\ref{eq:stepGdiag}), (\ref{eq:stepGr})
and (\ref{eq:stepGl}). \\ 
(3) New collision integrals $I_1^>(T+\Delta,t), I_2^> (t,T+\Delta),I^\rceil (T+\Delta,-i\tau)$ 
and $I^\lceil (-i\tau, T+ \Delta)$ are calculated by inserting the new Green functions
for times $t,t' \leq T+\Delta$ into Eqs.(\ref{i1}) to (\ref{i4}).\\
(4) The values of the collision integrals and the Hartree-Fock self-energy are approximated
by $\bar{I}=(I(T)+I(T+\Delta))/2$ and $\bar{\Sigma}^{HF}=(\Sigma^{HF}(T)+ \Sigma^{HF} (T+\Delta))/2$
where $I(T)$ and $I(T+\Delta)$ are the collision terms calculated under points (1) and (3).\\
(5) The Green function is then propagated from $G(T)\rightarrow G(T+\Delta)$ 
using the average values $\bar{I}$ and $\bar{h}^{HF}=h(T+\Delta/2)+ \bar{\Sigma}^{HF}$
in Eqs.(\ref{eq:stepGgrt}), (\ref{eq:stepGless}), (\ref{eq:stepGdiag}), (\ref{eq:stepGr})
and (\ref{eq:stepGl}). \\
\\
This concludes the general time-stepping procedure for the Green functions.\\
We finally consider the calculation of $\tilde{W}^<$ and $\tilde{W}^\rceil$ from Eqs.~(\ref{w1}) and (\ref{w4}).
As a consequence of the symmetry relation (\ref{eq:Wsym1}), we only need to calculate $\tilde{W}^< (t,t')$ for $t < t'$.
In a time step from $T$ to $T+\Delta $ we need to calculate $\tilde{W}^< (t,T+\Delta)$ for $t\leq T+\Delta$
from the known values of $\tilde{W}^<(t,T)$ for $t \leq T$. 
The first term on the righthand side of Eq.(\ref{w1}) can be calculated
directly from $G^< (t,T+\Delta)$ and $G^> (T+\Delta,t)$. However, the last term $X^< (t,T+\Delta)$
of Eq.(\ref{w1}) depends on the, still undetermined, values $\tilde{W}^< (t,T+\Delta)$. We therefore employ an 
iterative scheme. As a first guess for $\tilde{W}^< (t,T+\Delta)$ we take 
$\tilde{W}^< (t,T+\Delta)=\tilde{W}^< (t,T)$ for $t \leq T$ and $\tilde{W}^< (T+\Delta,T+\Delta)=\tilde{W}^< (T,T)$.
We therefore use the values of $\tilde{W}^<$ on the known sides of the time square at
time $T$ (solid lines in Fig.\ref{fig:timesquare}) as initial guesses for the stepped sides 
(dotted lines in Fig.\ref{fig:timesquare}) at $T+\Delta$.
As an initial guess for the value of $\tilde{W}^<$ at the new diagonal point $(T+\Delta, T+ \Delta)$, 
we take the value at  the previous diagonal point $(T,T)$.
We then calculate the quantity $X^< (t,T+\Delta)$ for $t \leq T+\Delta$
and obtain a new value for $\tilde{W}^< (t,T+\Delta)$ from Eq.(\ref{w1}).
This value is then inserted back into the righthand side of Eq.(\ref{w1})
and the process is repeated until convergence is reached.
Similarly we initialize  $\tilde{W}^\rceil (T+\Delta,-i\tau)=\tilde{W}^\rceil (T,-i\tau)$
and solve Eq.(\ref{w4}) in the same manner as for $\tilde{W}^<$.\\
This concludes our derivation of the time-stepping algorithm of the KB equations.
The propagation method described here has been used in two recent Letters~\cite{dahlen07,myohanen08}
where also values for the numerical parameters are given. It is clear that the choice
of these parameters depends strongly on the type of system considered, and on the strength of
the applied external fields.

\section{Summary and conclusions}

We presented a detailed account of the KB equations
and discussed in detail their structure, initial conditions and symmetries. 
We developed an algorithm for the time-propagation of the
KB equations in which the symmetry relations for the Green functions
were used to reduce the set equations that needed to be solved.
In two recent Letters~\cite{dahlen07,myohanen08} we applied the method to the case
of atoms and molecules in external time-dependent fields and to the case of transient transport dynamics
of double quantum dots.
We therefore conclude that time-propagation of the KB equations can be used
as a practical method to calculate the nonequilibrium properties of a wide variety of
many-body quantum systems, ranging from
atoms and molecules to quantum dots and quantum wells.
Moreover, the present work can be readily extended to 
other Green function formalisms, such as the
Nambu formalism~\cite{Nambu:PR60,Schrieffer:book} for superconducting systems.
Also future extension to bosonic systems is straightforward.
Work along these lines is in progress.

\end{document}